\newif\ifarxiv
\arxivfalse 

\ifarxiv
\documentclass{jfm-arxiv}
\else
\documentclass[]{jfm}
\fi
\usepackage{graphicx}
\usepackage{newtxtext}
\usepackage{newtxmath}
\usepackage{natbib}
\usepackage{hyperref}
\usepackage{tikz}
\hypersetup{
    colorlinks = true,
    urlcolor   = blue,
    linkcolor  = blue,
    citecolor  = black,
}

\newcommand{\RomanNumeralCaps}[1]
\linenumbers

\usepackage{amsmath}
\usepackage{longtable}
\usepackage{enumitem}
\usepackage{siunitx}
\usepackage{physics}
\usepackage{caption}
\usepackage{graphicx}
\usepackage{bm}
\usepackage{float}

\definecolor{tabred}{HTML}{d62728}
\definecolor{tabblue}{HTML}{1f77b4}
\definecolor{taborange}{HTML}{ff7f0e}
\definecolor{tabgreen}{HTML}{2ca02c}
\definecolor{tabpurple}{HTML}{9467bd}
\definecolor{lightgreen}{HTML}{67b99a}
\definecolor{darkgreen}{HTML}{036666}
\definecolor{tabolive}{HTML}{bcbd22}

\newcommand{\purplediamond}{\tikz\draw[tabpurple, thick] (0,0.15) -- (0.15,0) -- (0,-0.15) -- (-0.15,0) -- cycle;} 
\newcommand{\blueplus}{\tikz[rotate around={0:(0,0)}]\draw[tabblue, thick] (-0.05,0.15) -- (0.05,0.15) -- (0.05,0.05) -- (0.15,0.05) -- (0.15,-0.05) -- (0.05,-0.05)--(0.05,-0.15) -- (-0.05,-0.15) -- (-0.05,-0.05) -- (-0.15,-0.05) -- (-0.15,0.05) -- (-0.05,0.05) -- cycle;} 
\newcommand{\greenplus}{\tikz[rotate around={0:(0,0)}]\draw[tabgreen, thick] (-0.05,0.15) -- (0.05,0.15) -- (0.05,0.05) -- (0.15,0.05) -- (0.15,-0.05) -- (0.05,-0.05)--(0.05,-0.15) -- (-0.05,-0.15) -- (-0.05,-0.05) -- (-0.15,-0.05) -- (-0.15,0.05) -- (-0.05,0.05) -- cycle;} 

\newcommand{\orangecross}{\tikz[rotate around={45:(0,0)}]\draw[taborange, thick] (-0.05,0.15) -- (0.05,0.15) -- (0.05,0.05) -- (0.15,0.05) -- (0.15,-0.05) -- (0.05,-0.05)--(0.05,-0.15) -- (-0.05,-0.15) -- (-0.05,-0.05) -- (-0.15,-0.05) -- (-0.15,0.05) -- (-0.05,0.05) -- cycle;} 

\newcommand{\bluestar}{%
\tikz\draw[tabblue, thick] (0,0.15) -- (0.05,0.05) -- (0.15,0.05) -- (0.075,-0.025) -- (0.1,-0.15) -- (0,-0.075) -- (-0.1,-0.15) -- (-0.075,-0.025) -- (-0.15,0.05) -- (-0.05,0.05) -- cycle;}

\newcommand{\purpletriangle}{\tikz\draw[tabpurple, thick] (0,0.15) -- (0.15,-0.15) -- (-0.15,-0.15) -- cycle;} 

\newcommand{\olivetriangle}{\tikz\fill[tabolive,thick] (0,0.15) -- (0.15,-0.15) -- (-0.15,-0.15) -- cycle;} 

\newcommand{\redstarfill}{%
\tikz \fill[tabred,thick] (0,0.15) -- (0.05,0.05) -- (0.15,0.05) -- (0.075,-0.025) -- (0.1,-0.15) -- (0,-0.075) -- (-0.1,-0.15) -- (-0.075,-0.025) -- (-0.15,0.05) -- (-0.05,0.05) -- cycle;}

\newcommand{\bluerottriangle}{\tikz[rotate around={180:(0,0)}]\draw[tabblue,thick] (0,0.15) -- (0.15,-0.15) -- (-0.15,-0.15) -- cycle;} 

\definecolor{M030}{HTML}{003554}
\definecolor{M228}{HTML}{006494}
\definecolor{M300}{HTML}{0582ca}
\definecolor{M400}{HTML}{00a6fb}

\def\mytitle{Scaling of wall pressure and the peak of streamwise turbulence intensity in compressible wall flows}
\def\myshorttitle{Scaling of wall pressure and the peak of streamwise turbulence intensity}
\ifarxiv
\title[\myshorttitle]{\vspace{-85pt}\mytitle}
\else
\title[\myshorttitle]{\mytitle}
\fi

\shortauthor{A.~M.~Hasan, P.~Costa, J.~Larsson and R.~Pecnik}

\author{Asif~Manzoor~Hasan\aff{1}
  \corresp{\email{a.m.hasan@tudelft.nl}}, 
  Pedro~Costa\aff{1},
  Johan~Larsson\aff{2} \and
  Rene~Pecnik\aff{1}
 \corresp{\email{r.pecnik@tudelft.nl}}
 }

\affiliation{\aff{1} Process \& Energy Department, Delft University of Technology, Leeghwaterstraat 39, 2628~CB, Delft, The Netherlands
\aff{2}Department of Mechanical Engineering,
University of Maryland,
College Park, MD~20742, USA}

\begin{document}

\newcommand{\lineMa}{\textcolor{M030}{\rule[1mm]{1.2cm}{0.4mm}}}
\newcommand{\lineMb}{\textcolor{M228}{\rule[1mm]{1.2cm}{0.4mm}}}
\newcommand{\lineMc}{\textcolor{M300}{\rule[1mm]{1.2cm}{0.4mm}}}
\newcommand{\lineMd}{\textcolor{M400}{\rule[1mm]{1.2cm}{0.4mm}}}

\newcommand{\changed}[1]{{\color{red}#1}}

\maketitle

\begin{abstract}
This paper develops scaling laws for wall-pressure root-mean-square (r.m.s.) and the peak of streamwise turbulence intensity, accounting for both variable-property and intrinsic compressibility effects---those associated with changes in fluid volume due to pressure variations. 
To develop such scaling laws, we express the target quantities as an expansion series in powers of an appropriately defined Mach number. The leading-order term is represented using 
the scaling relations developed for incompressible flows, 
but with an effective Reynolds number.
Higher-order terms capture intrinsic compressibility effects and are modeled as constant coefficients, calibrated using flow cases specifically designed to isolate these effects.
The resulting scaling relations are shown to be accurate for a wide range of turbulent channel flows and boundary layers.
\end{abstract}

\section{Introduction}\label{sec1}

Wall-pressure fluctuations significantly impact the structural integrity of surfaces as well as the noise they emit \citep{bull1996wall}. Their accurate prediction is vital for engineering applications, particularly in high-speed flows where such fluctuations become more intense and pose greater design challenges. 
As a result, the scaling behavior of wall-pressure fluctuations in compressible flows has been an active area of research for several decades \citep{laganelli1983wall, bernardini2011wall,ritos2019acoustic,Zhang_Wan_Liu_Sun_Lu_2022,gerolymos2023scaling,wan2024intrinsic}. 
In contrast, research aimed at understanding the scaling behavior of peak streamwise turbulence intensity is majorly focused on incompressible flows \citep[see, for example,][]{marusic2017scaling,chen2021reynolds,smits2021reynolds}.

In incompressible flows, neither wall pressure nor the peak of streamwise turbulence intensity collapse under wall scaling (i.e., scaled using the friction velocity $u_\tau$ and the viscous length scale $\delta_v$), but rather increase with the friction Reynolds number $Re_\tau$. 
Recently, with high Reynolds number experimental and numerical data, various semi-empirical scaling laws have been proposed to capture this increase with $Re_\tau$. There are particularly two schools of thought behind these scaling laws. One according to Townsend's attached eddy model, which advocates that the wall pressure and the peak streamwise turbulence intensity increase indefinitely as a logarithmic function of $Re_\tau$ \citep{marusic2017scaling,panton2017correlation,smits2021reynolds}. 
The other approach corresponds to the power-law theory developed by \cite{chen2021reynolds} which argues that, at infinitely high $Re_\tau$, both wall pressure and the peak intensity (among other quantities) should asymptote to a constant value. The power-law increase of the peak intensity was recently supported by the high Reynolds number pipe flow DNS ($Re_\tau \approx 12000$) of \cite{pirozzoli2024streamwise}.
While the debate between these two theories is still ongoing, the focus of this paper is to extend these scaling theories to variable-property and compressible flows, where other parameters such as the Mach number (for instance, the free-stream Mach number $M_\infty$) and the wall cooling parameter (for instance, the ratio $T_w/T_r$, where $T_w$ and $T_r$ correspond to the wall and adiabatic temperatures, respectively) also become important. 
(Note that other wall-cooling parameters---such as the diabatic parameter $\Theta = (T_w - T_\infty)/(T_r - T_\infty)$~\citep{zhang2014generalized, cogo2023assessment}, where $T_\infty$ is the free-stream temperature, and the Eckert number $Ec = (\gamma - 1) M_\infty^2 T_\infty / (T_r - T_w)$~\citep{wenzel2022influences}---have been found to be more effective in quantifying wall-cooling effects than $T_w/T_r$.)

\cite{kistler1963fluctuating} performed the first measurement of wall pressure fluctuations underneath supersonic boundary layers (free-stream Mach number $M_\infty\leq5$), followed by other experimental studies summarized in figure~1 of \cite{beresh2011fluctuating}. Based on such experimental datasets, \cite{laganelli1983wall} developed an engineering model for wall-pressure r.m.s. scaled by the free-stream dynamic pressure ($q_\infty = 0.5 \rho_\infty U_\infty^2$, where subscript `$\infty$' implies free-stream values).
The experimental measurements used to tune Laganelli's model were found to exhibit significant scatter, largely due to their high sensitivity to the measurement sensors \citep{beresh2011fluctuating}, thereby raising concerns about the model's accuracy. This high level of scatter also hindered the development of more accurate models \citep{beresh2011fluctuating}.

\cite{bernardini2011wall} reported one of the earliest wall-pressure r.m.s. data using direct numerical simulations (DNS). They found that for their supersonic adiabatic boundary layers ($T_w/T_r = 1$), wall-pressure r.m.s. scaled by wall shear stress $\tau_w$ (i.e., $p_{w,rms}^+$) collapses for data at similar $Re_\tau$. However, a strong Mach number effect was seen if $p_{w,rms}$ was scaled by $q_\infty$, suggesting $\tau_w$ to better characterize wall-pressure.
Similarly, \cite{duan2016pressure} observed a weak Mach number effect on $p_{w,rms}^+$ for their quasi-adiabatic ($T_w/T_r=0.76$) boundary layer at hypersonic Mach number ($M_\infty = 5.86$). However, \cite{zhang2017effect} observed that at the same $M_\infty$, $p_{w,rms}^+$ substantially increases when the wall is strongly cooled, i.e., $T_w/T_r = 0.25$. More recently, \cite{Zhang_Wan_Liu_Sun_Lu_2022} observed that $p_{w,rms}^+$ decreases with wall cooling at sub- and supersonic Mach numbers, but increases with wall cooling at hypersonic Mach numbers $M_\infty>5$. They subsequently re-tuned the constants in Laganelli's model to better fit their data. More recently, the same group \citep{wan2024intrinsic} proposed another scaling model for $p_{w,rms}^+$ as a function of the free-stream Mach number for adiabatic boundary layers. However, this model does not account for changes in $Re_\tau$ and $T_w/T_r$.

Like boundary layers, several DNS studies were performed to study wall pressure and its scaling in fully developed channel flows. 
\cite{yu2020compressibility} decomposed the pressure field into a rapid, slow, viscous and compressible part, and observed that the compressible pressure increases strongly with the bulk Mach number. Later, \cite{yu2022wall} observed that this increase is better characterized in terms of the friction Mach number $M_\tau$. 
More recently, \cite{gerolymos2023scaling}, using their comprehensive dataset of compressible channel flows, proposed a scaling relation for $p_{w,rms}^+$, which can be applied to both channels and boundary layers.

The wall-pressure scaling laws proposed for boundary layers, like the ones in \cite{laganelli1983wall} and \cite{wan2024intrinsic}, have not been generalized to other class of flows, like channel and pipe flows. 
Moreover, the model of \cite{gerolymos2023scaling}, which was originally tested for channel flows, leads to significant errors for high-Mach-number boundary layers. Clearly, a universally applicable and accurate scaling law is lacking.  

Compared to wall-pressure fluctuations, less work has been done to study the scaling behavior of peak streamwise turbulence intensity in compressible flows.
For variable-property channel flow cases at zero Mach number, \cite{patel2015semi} observed that the peak of streamwise turbulence intensity can be higher or lower than a corresponding incompresible flow at similar $Re_\tau$, depending on the distribution of the semi-local Reynolds number $Re_\tau^*$ (defined as $\bar \rho u_\tau^* h /\bar \mu$, where $\bar \rho$ and $\bar \mu$ imply mean density and viscosity, and $u_\tau^* = \sqrt{\tau_w/\bar \rho}$ is the semi-local friction velocity). Specifically, the peak intensity is higher if $Re_\tau^*$ decreases away from the wall and lower if it increases. 

For flows at non-zero Mach numbers, the peak value is found to be higher than a corresponding incompressible flow at similar $Re_\tau$, independent of the distribution of $Re_\tau^*$ \citep{gatski2002numerical,pirozzoli2004direct,foysi2004compressibility,duan2010direct,modesti2016reynolds,zhang2018direct,trettel2019transformations,cogo2022direct,cogo2023assessment}. In \cite{hasan2024intrinsic}, it was concluded that the higher value of the peak is due to intrinsic compressibility effects\textemdash those associated with changes in fluid volume in response to changes in pressure \citep{lele1994compressibility}. However, a formal scaling law which accounts for these effects is missing.

In this paper, we develop scaling laws for wall-pressure r.m.s. and the peak of streamwise turbulence intensity that account for compressibility effects\textemdash variable-property and intrinsic compressibility \citep{lele1994compressibility,hasan2024intrinsic}\textemdash and are applicable to both channel/pipe flows and boundary layers.
To develop such scaling laws, we express wall-pressure r.m.s. and the peak intensity as an expansion series in powers of an appropriately defined Mach number \citep{ristorcelli1997pseudo}.  
The leading-order term in this series accounts for Reynolds number and variable-property effects, and is represented by using the same scaling laws as developed for incompressible flows \citep{chen2022law}, however, with an effective value of the semi-local friction Reynolds number, instead of the wall-based $Re_\tau$. 
The higher-order terms mainly account for intrinsic compressibility effects, and are modeled using the constant-property high-Mach-number cases of \cite{hasan2024intrinsic}, which are designed to isolate intrinsic compressibility effects.

\section{Approach}\label{sec2}

\cite{ristorcelli1997pseudo} expressed the compressible flow quantities such as pressure, density, and velocity in the form of an expansion series as follows:
\begin{equation}\label{prerist}
    \begin{aligned}
    p^{\prime} / \mathcal{P} &= \epsilon^2 \left[ p_1^{\prime} / (\rho_\infty \mathcal{U}^2) + \epsilon^2\, p_2^{\prime} / (\rho_\infty \mathcal{U}^2) + \ldots \right], \\
    \rho^{\prime} / \rho_\infty &= \epsilon^2 \left[ \rho_1^{\prime} / \rho_\infty + \epsilon^2\, \rho_2^{\prime} / \rho_\infty + \ldots \right], \\
    u_i / \mathcal{U} &= u_{0i} / \mathcal{U} + \epsilon^2 \left[ u_{1i} / \mathcal{U} + \epsilon^2\, u_{2i} / \mathcal{U} + \ldots \right].
    \end{aligned}
\end{equation}
where $\mathcal P$ is the thermodynamic scale of pressure, $\mathcal U$ is the velocity scale, $\rho_\infty$ is the reference background density,
$\epsilon = \rho_\infty \mathcal U^2/\mathcal P$ is the ratio of the hydrodynamic scale of pressure $\rho_\infty \mathcal U^2$ to the thermodynamic scale, and the subscripts `0, 1, 2, 3' signify the order of the variables. These expansions, when substituted in the Navier-Stokes equations and the terms with similar powers of $\epsilon$ are grouped, we get a set of zeroth order (proportional to $\epsilon^0$), first order (proportional to $\epsilon^2$) and higher order governing equations. The zeroth order set of equations solves for the incompressible velocity $u_{0i}$ and pressure $p_1$\footnote{Note that, since the pressure term in the Navier-Stokes equation, upon appropriate non-dimensionalization, is multiplied with $\epsilon^{-2}$, a first order pressure $p_1$ acts as the incompressible pressure to satisfy a meaningful balance between velocity and pressure terms \citep{ristorcelli1997pseudo}.}, whereas the higher order equations solve for the higher order variables.

It is worth noting that, in these expansions, \cite{ristorcelli1997pseudo} assumed the thermodynamic scale of pressure to be the reference background pressure $P_\infty$, and the velocity scale to be $(2 k/3)^{1/2}$, where $k$ is the turbulent kinetic energy defined as $k = \overline{u_i^{\prime\prime}u_i^{\prime\prime}}/2$. This gives $\epsilon^2 =  \rho_\infty (2k/3)/P_\infty$, which is equal to $\gamma M_t^2$ for ideal gas flows, where $\gamma$ is the ratio of specific heats and $M_t =(2 k/3)^{1/2}/c_\infty$, with $c_\infty$ being the reference speed of sound.

Let us now extend Ristorcelli's approach to compressible wall-bounded flows, however, with some notable differences. 
For homogeneous flows, \cite{ristorcelli1997pseudo} represented the compressible flow field as a sum of an incompressible field and higher order fields, with the latter capturing effects that arise only at finite Mach numbers. In the same spirit, we represent a compressible wall-bounded flow field as a sum of a variable-property zero-Mach-number field\textemdash having the same Reynolds number and mean property distributions as the compressible flow\textemdash and higher order fields representative of finite-Mach-number effects. Such an expansion ensures that the zeroth-order term in the expansion series comprises of variable-property and Reynolds number effects, such that the higher order terms primarily represent intrinsic compressibility effects which occur only at finite Mach numbers. 

Some other differences are listed as follows. 
Since the base (zeroth order) state about which the expansion series is built, represents a zero-Mach-number flow with mean property variations, the reference density $\rho_\infty$ and $c_\infty$ in Ristorcelli's work should be replaced with the mean density $\bar \rho$ and the mean speed of sound $\bar c$, respectively. 
Additionally, the semi-local friction velocity scale $u_\tau^*$ becomes the relevant scale in compressible wall-bounded flows, instead of $(2k/3)^{1/2}$. 
Lastly, we choose the thermodynamic pressure scale to be $\bar \rho \bar c^2$ rather than $\bar p$, for the following reason. The isentropic density fluctuations are related to pressure fluctuations through the relation
\(
    {({\rho^\prime})^{is}}/{\bar\rho} \approx {p^{\prime}}/({\bar \rho \bar c^2})
\) \citep{hasan2024intrinsic}.
This implies that the density fluctuations depend on $\bar \rho \bar c^2$.
Since intrinsic compressibility effects are, by definition, related to isentropic density fluctuations, it is natural that the parameter characterizing these effects\textemdash namely, $\epsilon$\textemdash also depends on $\bar \rho \bar c^2$.
Taking $\mathcal{P}= \bar \rho \bar c^2$, along with $\bar \rho {u_\tau^*}^2$ as the hydrodynamic pressure scale, we get
\begin{equation}\label{eps}
    \epsilon^2 = {\bar \rho {u_\tau^*}^2}/({\bar \rho \bar c^2}) = {M_\tau^*}^2.
\end{equation} 

By accounting for these differences, we get the expansion series for pressure fluctuations in wall-bounded flows (analogous to equation~\ref{prerist} for homogeneous flows) as
\begin{equation}\label{pt}
    p^{\prime} / (\bar{\rho} \bar{c}^2) = {M_\tau^*}^2 \left[ p_1^{\prime} / (\bar{\rho} {u_\tau^*}^2) + {M_\tau^*}^2\, p_2^{\prime} / (\bar{\rho} {u_\tau^*}^2) + \ldots \right].
\end{equation}

By writing this equation at the wall, squaring it, averaging, and dividing by ${M_\tau^*}^4$, we get the equation for wall-pressure variance, scaled by $\tau_w^2$ as
\begin{equation}\label{pp}
    {\overline{p^{\prime}p^{\prime}}}_w^+ = \overline{p^{\prime}_1p^{\prime}_1}_w^+ + {M_\tau^*}^2 \left[ \overline{ 2 p^{\prime}_1p^{\prime}_2}_w^+ + {M_\tau^*}^2 (\overline{p^{\prime}_2 p^{\prime}_2}_w^+ +\overline{2 p^{\prime}_1 p^{\prime}_3}_w^+) + \ldots\right],
\end{equation}
where the first term on the right-hand-side signifies wall-pressure variance in a zero-Mach-number variable-property flow.

At this point, it is important to note that not only the leading-order correlation, but also other higher-order correlations on the right-hand side are mainly influenced by Reynolds number and variable-property effects. This is justified as follows.
From the analysis of \cite{ristorcelli1997pseudo}, we note that the first-order equations\textemdash governing the evolution of first-order velocity, second-order density, and second-order pressure ($u_{i1}$, $\rho_2$, and $p_2$)\textemdash depend explicitly on the leading-order (incompressible) variables: $u_{i0}$, $\rho_1$, and $p_1$.
For wall-bounded flows, this implies that these higher-order variables ($u_{i1}$, $\rho_2$, $p_2$) are indirectly affected by Reynolds number and variable-property effects, through their dependence on the incompressible solution.
Similarly, even higher-order quantities, such as $u_{i2}$, $\rho_3$, $p_3$, are primarily influenced by these effects through their dependence on lower-order variables.
These higher-order quantities are not explicitly affected by intrinsic compressibility effects since there is no Mach number or $\epsilon$ in the set of governing equations \citep{ristorcelli1997pseudo}; instead, such effects are embedded in the parameter $\epsilon$, by which these variables are multiplied in the expansion series.

Given this understanding, we model the correlations in equation~\eqref{pp} as a sum of a constant and a function which depends on Reynolds number and variable-property effects, inspired from the relations proposed for incompressible flows \citep{chen2022law}. 
For instance, 
\(
    \overline{p^{\prime}_1p^{\prime}_1}_w^+ = c_{0,p} +f_{0,p}, \)
where $c_{0,p}$ is a constant, $f_{0,p}$ is an unknown function, and the subscript `$0,p$' signifies the leading order term for pressure. Representing all other higher order correlations also in a similar form and substituting them in equation~\eqref{pp}, we get 
\begin{equation}\label{ppc}
    {\overline{p^{\prime}p^{\prime}}}_w^+ = \underbrace{c_{0,p}+f_{0,p}}_{\textrm{$Re$ \& VP}} + \underbrace{{M_\tau^*}^2 c_{1,p} + {M_\tau^*}^4 c_{2,p} + \ldots}_{\textrm{IC}} + \underbrace{{M_\tau^*}^2 f_{1,p}+ {M_\tau^*}^4 f_{2,p}+ \ldots}_{\textrm{coupling $Re$, VP, IC}},
\end{equation}
where `$Re$' denotes contribution by Reynolds number effects, `VP' variable-property effects and `IC' intrinsic compressibility effects. The right-most expression under the brace highlights coupling between these effects.

Here, we postulate that the coupling effects are small, and can be neglected. We test this hypothesis aposteriori based on the available data. 
From this simplification, we get
\begin{equation}\label{ppc2}
    {\overline{p^{\prime}p^{\prime}}}_w^+ \approx {c_{0,p}+f_{0,p}} + {M_\tau^*}^2 c_{1,p} + {M_\tau^*}^4 c_{2,p} + \ldots.
\end{equation}

Following a similar approach for the inner-scaled peak streamwise turbulence intensity, namely, ${\widetilde{u^{\prime\prime}u^{\prime\prime}}}_p^* = \overline{ \rho u^{\prime\prime} u^{\prime\prime}}_p/\tau_w$ (where $\widetilde{.}$ represents Favre averaging), we get   
\begin{equation}\label{uuc}
    {\widetilde{u^{\prime\prime}u^{\prime\prime}}}_p^* \approx {c_{0,u}+f_{0,u}}+ {M_\tau^*}^2 c_{1,u} + {M_\tau^*}^4 c_{2,u} + \ldots,
\end{equation}
where $c_{0,u}$, $c_{1,u}$, etc. are constants, analogous to $c_{0,p}$, $c_{1,p}$ for wall-pressure above. The first term on the right-hand-side represents the leading order correlation ${\widetilde{u_0^{\prime\prime}u_0^{\prime\prime}}}_p^*$.

Before proceeding, we note that for ideal gases, the semi-local friction Mach number is approximately uniform \citep{hasan2024intrinsic}. This also holds for the non-ideal gas cases analysed here \citep{sciacovelli2017direct}. Thus, hereafter, we assume
\(
    M_\tau^* \approx M_\tau.
\)

In the following subsections, we will model the unknown functions $f_{0,p}$ and $f_{0,u}$, and determine the constants in equations~\eqref{ppc2}~and~\eqref{uuc}.

\subsection{Variable-property effects}\label{sec3}

\cite{patel2015semi} showed that both semi-locally-scaled wall-pressure and the peak streamwise intensity (along with other quantities) are similar for flows with similar distributions of $Re_\tau^*$, independent of the distribution of $\bar \rho$ and $\bar \mu$. Through this they confirmed that variable-property effects simply change the local Reynolds number of the flow, conjectured earlier in \cite{morkovin1962effects} and \cite{spina1994physics}. 
Consequently, a natural choice to model $\overline{p^{\prime}_1p^{\prime}_1}_w^+ = c_{0,p}+f_{0,p}$ and ${\widetilde{u_0^{\prime\prime}u_0^{\prime\prime}}}_p^* = c_{0,u}+f_{0,u}$ would be to use the scaling relations developed for incompressible flows \citep{chen2022law}, with an effective value of $Re_\tau^*$.

It is well established in the incompressible flow literature that the dominant contribution to wall pressure arises from the source terms in the buffer layer \citep{kim1989structure,kim1993propagation}. In light of this, we propose that $Re_\tau^*$ computed in the buffer layer, say at $y^*=15$, should be used for scaling wall-pressure.

\begin{figure}
    \centering\includegraphics[width = 1\textwidth]{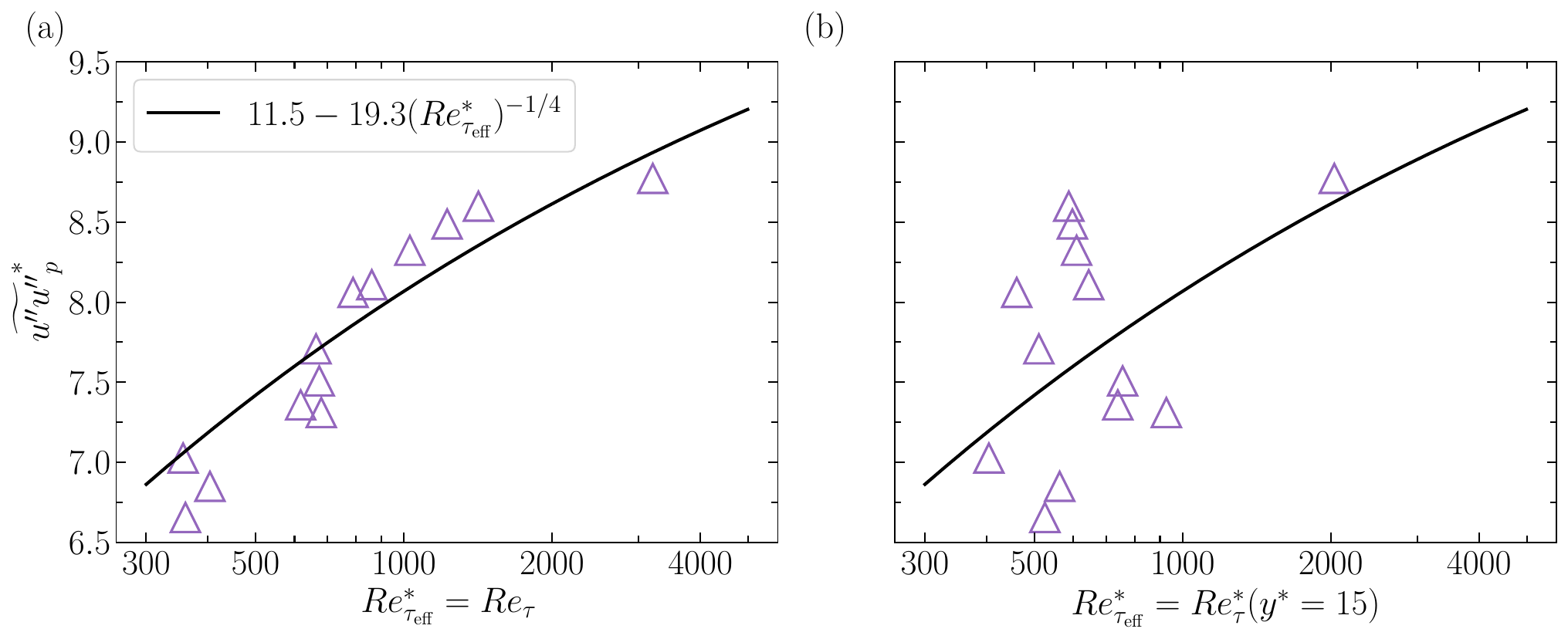} 
\caption{Semi-locally scaled streamwise turbulent peak intensity, i.e., ${\widetilde{u^{\prime\prime}u^{\prime\prime}}}_p^* = \overline{\rho u^{\prime\prime}u^{\prime\prime}}_p/\tau_w$ as a function of (a) $Re_\tau$ and (b) $Re_\tau^*$ taken at the peak location ($y^*=15$), for the low-Mach-number variable-property cases of \cite{modesti2024friction}.
The black curve corresponds to the fit proposed in \cite{chen2022law} for incompressible flows.} 
\label{uuinc}
\end{figure}
\begin{table}
    \centering 
    \begin{tabular}{m{2.4cm} >{\centering\arraybackslash}m{0.8cm} >{\centering\arraybackslash}m{7.5cm} >{\centering\arraybackslash}m{0.8cm} >{\centering\arraybackslash}m{0.8cm}}
    Quantity $\phi$ & $c_{0,\phi}$&  $f_{0,\phi}$& $c_{1,\phi}$& $c_{2,\phi}$\\ \hline
    {\textbf{Channels \& Pipes}}&\\ 
   $ {\overline{p^{\prime}p^{\prime}}}_w^+$  & 19.36 & $- 92.4 \big[Re_\tau^*(y^*=15)\big]^{-1/4} + 110.25 \big[Re_\tau^*(y^*=15)\big]^{-1/2}$     & 2.4 & 8312.5\\
   ${\widetilde{u^{\prime\prime}u^{\prime\prime}}}_p^*$  & 11.5 & $-19.3 Re_\tau^{-1/4}$     & 78.9 & 199.3\\
   {\textbf{Boundary layers}}&\\    
   $ {\overline{p^{\prime}p^{\prime}}}_w^+$ & 20.25 & $-87.3 \big[Re_\tau^*(y^*=15)\big]^{-1/4} + 94.09 \big[Re_\tau^*(y^*=15)\big]^{-1/2}$     & 2.4 & 8312.5\\ 
   ${\widetilde{u^{\prime\prime}u^{\prime\prime}}}_p^*$  & 11.5 & $-19.3Re_\tau^{-1/4}$     & 78.9 & 199.3\\
\end{tabular}
    \captionof{table}{The constants and functions in equations~\eqref{ppc2}~and~\eqref{uuc}, i.e., $\phi = c_{0,\phi} + f_{0,\phi} + c_{1,\phi} M_\tau^2 + c_{2,\phi} M_\tau^4$ (neglecting higher order terms). The constant $c_{0,\phi}$ and the function $f_{0,\phi}$ are taken from \cite{chen2022law}, as discussed in section~\ref{sec3}. The constants $c_{1,\phi}$ and $c_{2,\phi}$ are calibrated based on the constant-property high-Mach-number cases \citep{hasan2024intrinsic}, as discussed in section~\ref{sec4}.}
    \label{tabconst}
\end{table}

For the peak intensity, the choice is not as straightforward as it was for wall-pressure r.m.s. One obvious choice is to compute the Reynolds number at the peak location itself ($y^*\approx15$). Another choice is to use the wall $Re_\tau$. The motivation for the latter choice comes from the analysis of \cite{bradshaw1967inactive}, who argued that the increase in the peak intensity with $Re_\tau$ for incompressible flows is directly associated with the large-scale fluctuations in wall shear stress. This is also mathematically supported by the Taylor series expansion of $\overline{u^\prime u^\prime}^+$ at the wall \citep{chen2021reynolds,smits2021reynolds}, where the leading order term represents fluctuations in wall shear stress. 

To assess the correct Reynolds number that accounts for variable-property effects on wall-shear-stress fluctuations, and hence, the peak intensity, we analyse the low-Mach-number variable-propety cases of \cite{modesti2024friction}. These cases are essentially free of intrinsic compressibility effects and therefore quantify variable-property effects.
For these cases, we have computed the wall shear stress fluctuations using the interpolation technique in \cite{smits2021reynolds}, and we observed that these fluctuations scale with $Re_\tau$, and approximately follow the relation in \cite{chen2021reynolds}, i.e., $0.25-0.42 Re_\tau^{-1/4}$ (not shown). Following the discussion presented above, this implies that the peak intensity should also scale with $Re_\tau$. This is verified in figure~\ref{uuinc}, which shows the semi-locally scaled peak intensity ${\widetilde{u^{\prime\prime}u^{\prime\prime}}}_p^*$ as a function of (a) $Re_\tau$ and (b) $Re_\tau^*$ taken at the peak location, i.e., $Re_\tau^*(y^*=15)$. Clearly, the spread in the data with respect to the fit from \cite{chen2022law} is lower for $Re_\tau$ than for $Re_\tau^*(y^*=15)$.
Note that since these cases are at negligible Mach numbers, their peak intensity is a direct measure of the leading order term in the expansion series, i.e., ${\widetilde{u_0^{\prime\prime}u_0^{\prime\prime}}}_p^*$.

Even though $Re_\tau$ is a better choice than the Reynolds number at the peak location, there is still some spread in the data around the curve fit (see figure~\ref{uuinc}a). This spread is mainly attributed to the effects associated with the gradients in the semi-local Reynolds number \citep{patel2015semi}, which will be neglected here.

Finally, with these choices of the Reynolds numbers, we model the leading order terms in equations~\eqref{ppc2}~and~\eqref{uuc} as described in table~\ref{tabconst}.

\subsection{Intrinsic compressibility effects}\label{sec4}

We now determine the higher order constants in equations~\eqref{ppc2}~and~\eqref{uuc} using the constant-property high-Mach-number cases of \cite{hasan2024intrinsic}, designed to isolate intrinsic compressibility effects.

\begin{figure}
    \centering\includegraphics[width = 1\textwidth]{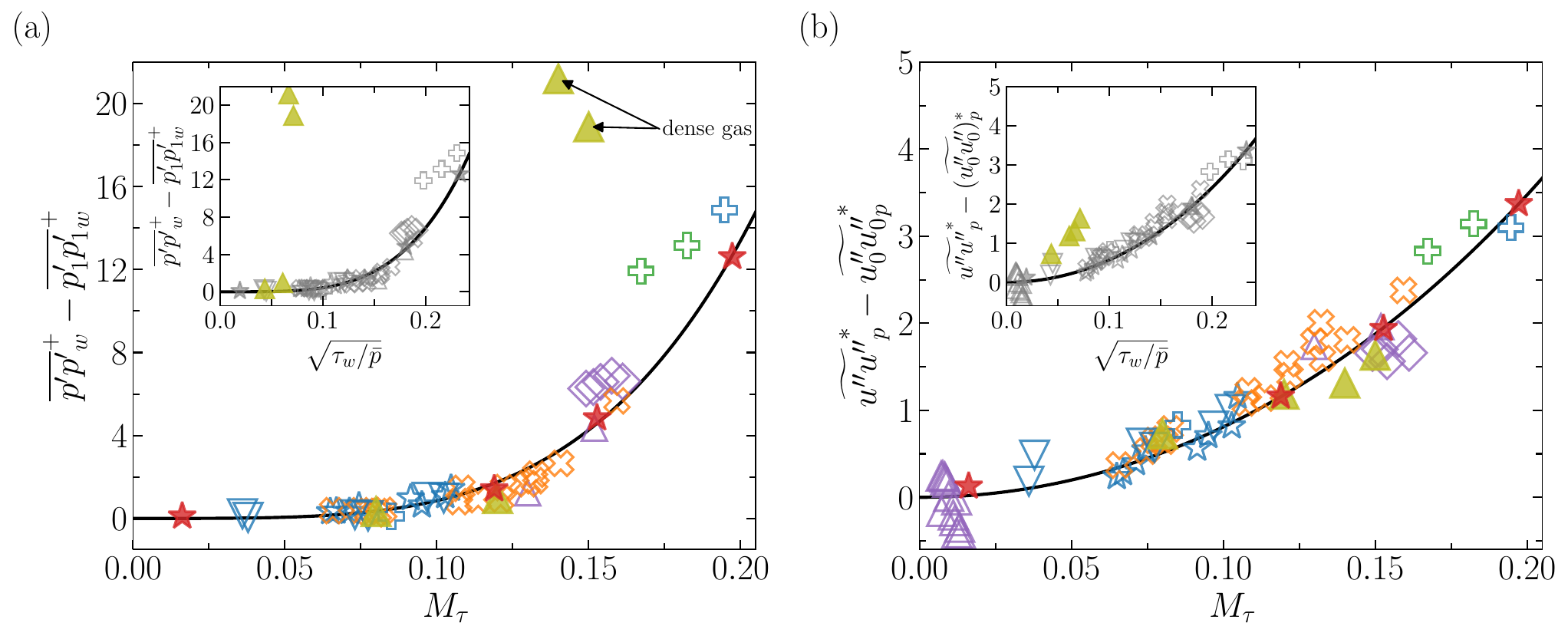} 
\caption{Contribution by intrinsic compressibility effects to (a) wall-pressure variance and (b) the peak streamwise turbulence intensity as a function of $M_\tau$ for a wide range of channels and boundary layers taken from the following studies. 
\textbf{Channels:} 
\scalebox{0.7}{\protect\bluerottriangle}$\times 6$-\cite{trettel2016mean},
\scalebox{0.8}{\protect\olivetriangle}$\times 4$-\cite{sciacovelli2017direct},
\scalebox{0.7}{\protect\purpletriangle}$\times 13$-\cite{modesti2024friction},
\scalebox{0.8}{\protect\redstarfill}$\times 4$-\cite{hasan2024intrinsic};
\textbf{Boundary layers:} \scalebox{0.7}{\protect\bluestar}$\times 8$-\cite{bernardini2011wall},
\scalebox{0.7}{\protect\blueplus}$\times 2$-\cite{zhang2018direct},
\scalebox{0.7}{\protect\purplediamond}$\times 5$-\cite{ceci2022numerical},
\scalebox{0.7}{\protect\greenplus}$\times 2$-\cite{huang2022direct},
\scalebox{0.7}{\protect\orangecross}$\times 16$-\cite{cogo2022direct,cogo2023assessment},
\scalebox{0.7}{\protect\purpletriangle}$\times 2$-A.~Ceci (private communication).
 The black curve in (a)~and~(b) corresponds to $c_{1,\phi} M_\tau^2 + c_{2,\phi} M_\tau^4$, where $c_{1,\phi}$ and $c_{2,\phi}$ are reported in table~\ref{tabconst}, and $\phi$ represents wall-pressure r.m.s. or the peak intensity.
(Insets) Intrinsic compressibility contribution as a function of $\sqrt{\tau_w/\bar p}$. The gray symbols signify ideal gas air cases for which $\sqrt{\tau_w/\bar p} = \sqrt{1.4} M_\tau$, and the colored symbols represent the dense gas cases of \cite{sciacovelli2017direct}. Note that low-Reynolds-number cases are excluded.}
\label{icpciu}
\end{figure}

Let us first focus on wall-pressure r.m.s.
    To obtain the constants, we first substract the variable-property contribution~($c_{0,\phi}+f_{0,\phi}$; see table~\ref{tabconst}) from the total $\overline{p^{\prime}p^{\prime}}_w^+$ taken from the DNS. Next, we plot this difference\textemdash which signifes the contribution by intrinsic compressibility effects\textemdash as a function of the expansion parameter $\epsilon = M_\tau$ in figure~\ref{icpciu}(a) for the four constant-property cases (denoted by red stars). Fitting a curve of the form $ M_\tau^2 c_{1,p} +  M_\tau^4 c_{2,p}$ (neglecting higher order terms) to these cases, we obtain $c_{1,p} = 2.4$ and $ c_{2,p} =8312.5$; see figure~\ref{icpciu}(a)~(black curve).

Figure~\ref{icpciu}(a) plots the intrinsic compressibility contribution for several boundary layer and channel flow cases in the literature as listed in the caption.
Despite the fact that these cases are at different Reynolds numbers, and possess different distributions of mean properties, majority of the cases follow the curve fit set by the constant-property cases quite well. This corroborates the assumption we made regarding neglecting the coupling terms in section~\ref{sec2}. 
However, there are some exceptions. For the boundary layer cases of \cite{huang2022direct}, represented by green plus, the deviation from the curve fit is quite high. This could be due to the simplications we made in the analysis, or due to issues with obtaining a converged pressure profile in DNS. Also, the dense gas cases of \cite{sciacovelli2017direct} at high Mach numbers depict extremely high wall-pressure fluctuations. This is due to the proximity of these cases to the Widom line (the curve that marks the maximum of the specific heat at constant pressure, above the critical point). In this region, small pressure fluctuations can cause large density fluctuations, which in turn intensify pressure fluctuations through the source terms. Such an effect is
due to the complexity of the equation of state, and is thus not accounted for in the present scaling model. 

Repeating the same analysis for the peak of streamwise turbulence intensity, we get
$c_{1,u} = 78.9$ and $ c_{2,u} = 199.3$.
Figure~\ref{icpciu}(b) shows the intrinsic compressibility contribution towards the peak intensity for the constant- and variable-property cases listed in the caption, along with the curve fit with tuned constants.
Clearly, majority of the cases follow the curve, corroborating that the coupling effects are small.

The insets in figure~\ref{icpciu} show the intrinsic compressibility contributions to the wall-pressure r.m.s. and the peak intensity as functions of $\sqrt{\tau_w/\bar p}$\textemdash the form that $\epsilon$ would take if $\bar p$ were chosen as the thermodynamic pressure scale.
For the ideal gas air cases (shown in gray symbols), $\sqrt{\tau_w/\bar p}$ quantifies intrinsic compressibility effects as effectively as $M_\tau$ (main figure~\ref{icpciu}, discussed earlier), since $\sqrt{\tau_w/\bar p} \approx \gamma M_\tau$, with $\gamma=1.4$ for these cases. 
However, for the dense gas cases of \cite{sciacovelli2017direct} (shown in colored symbols), the characterization of intrinsic compressibility effects deteriorates for both wall-pressure and the peak when $\sqrt{\tau_w/\bar p}$ is used instead of $M_\tau$. 
These observations support our choice made in section~\ref{sec2} of using $\bar \rho \bar c^2$ as the relevant thermodynamic scale of pressure rather than $\bar p$. This is further supported by the observation that, for the dense gas cases, the upward shift in the logarithmic mean velocity profiles due to intrinsic compressibility effects \citep{hasan2023incorporating} is well quantified in terms of $M_\tau$, instead of $\sqrt{\tau_w/\bar p}$ (not shown).

\textbf{The final scaling relation}\textemdash 
By combining the functions and constants obtained from sections~\ref{sec3}~and~\ref{sec4}, we get the final scaling relations for the wall-pressure r.m.s. and the peak streamwise turbulence intensity, as reported in table~\ref{tabconst}.
Using these relations, we compute errors with respect to the DNS as
\(
    \varepsilon_{\phi} = ({\phi^{DNS} - \phi^{model}})/{\phi^{DNS}},
\)
where $\phi$ represents the wall-pressure r.m.s. or the peak intensity.
We also compute the root-mean-square (RMS) error across all cases as
\(
    \textrm{RMS} = \sqrt{\frac{1}{N} \sum \varepsilon_\phi^2},
\)
where $N$ is the total number of cases.

For wall-pressure r.m.s., the maximum absolute error and RMS error across all the cases presented in figure~\ref{icpciu} are 15.6\% and 4.6\%, respectively, whereas for the peak intensity, the corresponding values are 6.1\% and 2.8\%. (Note that the dense gas cases were excluded from the wall-pressure error calculations.) 
Moreover, when comparing the RMS errors obtained using different Reynolds number definitions in \( f_{0,\phi} \)---namely \( Re_\tau \), \( Re_\tau^*(y^* = 15) \), \( Re_\tau^*(y^* = 50) \), \( Re_\tau^*(y/\delta = 0.2) \), and \( Re_\tau^*(y/\delta = 1) \)---we find that the definitions adopted in section~\ref{sec3} yield relatively lower RMS values than those associated with the other choices.

\begin{table}
    \centering 
    \begin{tabular}{m{2.5cm} >{\centering\arraybackslash}m{2.2cm} >{\centering\arraybackslash}m{1.6cm} >{\centering\arraybackslash}m{1.8cm} >{\centering\arraybackslash}m{1.2cm} >{\centering\arraybackslash}m{1.5cm} >{\centering\arraybackslash}m{1.2cm}}
    & \cite{laganelli1983wall}$^\dagger$&  \cite{ritos2019acoustic}$^\dagger$ &\cite{Zhang_Wan_Liu_Sun_Lu_2022}$^\dagger$& G\&V (2023)$^\ddagger$ & \cite{wan2024intrinsic}$^\dagger$ & Present$^\star$\\ \hline 
    Max abs error[\%] & 39 & 34.3    & 41.9 & 15.7 &18.3 & 15.6 \\ 
    RMS[\%]   & 29    & 12.4    & 13.3&  6.9 &9.5 &4.6\\ 
    \end{tabular}
    \captionof{table}{The maximum absolute error ($L_\infty$ norm) and the RMS error ($L_2$ norm; defined in the main text) for various wall-pressure r.m.s. models available in the literature. The models marked with `$\dagger$'~have been applied exclusively to conventional (ideal-gas air) boundary layers, for which they were originally developed. The model marked with `$\ddagger$'~has been tested on both conventional channels and boundary layers. Finally, the present model marked by `$\star$' has been applied to a broader set of flows, including conventional channels and boundary layers, as well as the four constant-property cases reported in \cite{hasan2024intrinsic}.
    Note that G\&V stands for \cite{gerolymos2023scaling}.}
    \label{prmstab}
    \end{table}
    
\textbf{Comparison with existing models}\textemdash 
Table~\ref{prmstab} reports the maximum absolute error and RMS error in predicting the wall-pressure r.m.s. using various models from the literature.  
As seen, the Laganelli family of models \citep{laganelli1983wall,ritos2019acoustic,Zhang_Wan_Liu_Sun_Lu_2022} yield relatively high error values, with an RMS error exceeding 12\%. The model proposed by \cite{wan2024intrinsic} achieves better accuracy, with an RMS error of 9.5\%, despite not explicitly accounting for Reynolds number or wall-cooling effects. Among the models available in the literature, the best performance is observed for the model by \cite{gerolymos2023scaling}, which achieves an RMS error of 6.9\%. However, this value is still high, primarily due to significant errors incurred in predicting high-Mach-number boundary layers.
The present model shows improved performance over existing approaches, with the lowest RMS error of 4.6\%.

A similar comparative analysis for the peak of streamwise turbulence intensity could not be conducted, as, to the best of our knowledge, there are no other models for this quantity in compressible flows.

\section{Summary}\label{sec6}
In this paper, we have proposed scaling relations for the wall-pressure root-mean-square (r.m.s.) and the peak of streamwise turbulence intensity that are applicable to both channels and boundary layers. These relations were developed by expressing these quantities as an expansion series in terms of the friction Mach number, $M_\tau$. 
The first term in this series accounts for Reynolds number and variable-property effects, whereas the higher-order terms primarily capture intrinsic compressibility effects. 

To model the leading order term, we used the same expressions as proposed for incompressible flows, with an effective value of the semi-local Reynolds number which incorporates variable-property effects. For wall-pressure r.m.s, this value was found to be the Reynolds number defined in the buffer layer (at $y^*=15$), whereas for the peak streamwise intensity, we found the effective value to be the wall $Re_\tau$.

We model the higher-order correlations in the series as constants, whose values are found based on our constant-property high-Mach-number cases representative of intrinsic compressibility effects \citep{hasan2024intrinsic}. Modeling these correlations as simple constants implies that any coupling between Reynolds number, variable-property and intrinsic compressibility effects is small\textemdash an assumption which was verified aposteriori using the available data.

Finally, an additional key finding has been noted: based on the dense gas (non-ideal) cases of \cite{sciacovelli2017direct}, we confirm that $\bar\rho \bar c^2$ is a more appropriate thermodynamic pressure scale than the mean pressure $\bar p$, supporting the choice of expressing the expansion series in terms of $M_\tau$.
Future work should extend the proposed scaling approach to other turbulence quantities, and explore the role of higher-order terms and coupling effects relevant at Mach numbers beyond those considered here.

This work was supported by the European Research Council grant no.~ERC-2019-CoG-864660, Critical; and the Air Force Office of Scientific Research under grant FA9550-23-1-0228. We thank Sergio Pirozzoli (University of Rome) for the insightful discussions. 

\backsection[Declaration of Interests]{The authors report no conflict of interest.}


\end{document}